\documentclass[11pt]{article}
\pdfoutput=1
\usepackage[centertags]{amsmath}
\usepackage[square, comma, sort&compress,numbers]{natbib}
\usepackage{array,multirow}
\numberwithin{equation}{section}
\usepackage{amssymb,mathtools}
\usepackage{amsfonts,amsthm,stmaryrd,bm}
\usepackage{graphicx}
\usepackage{color}
\usepackage[normalem]{ulem}

\usepackage{epsfig}
\usepackage{bbold}

\usepackage{wrapfig}
\usepackage{float}
\usepackage{soul}
\usepackage{color}
\usepackage{enumitem}

\usepackage[dvipsnames]{xcolor}

\usepackage{tikz,pgf}
\usetikzlibrary{shapes}
\usetikzlibrary{calc}
\usetikzlibrary{decorations.pathmorphing}
\usetikzlibrary{decorations.pathreplacing,shapes.misc}
\usetikzlibrary{positioning}
\usetikzlibrary{arrows}
\usetikzlibrary{decorations.markings}
\usetikzlibrary{shadings}

\usetikzlibrary{intersections}

\def\sideremark#1{\ifvmode\leavevmode\fi\vadjust{\vbox to0pt{\vss
 \hbox to 0pt{\hskip\hsize\hskip1em
 \vbox{\hsize3cm\tiny\raggedright\pretolerance10000
  \noindent #1\hfill}\hss}\vbox to8pt{\vfil}\vss}}}

\def\dps{\displaystyle}

\def\tr{{\rm Tr}}

\renewcommand{\tilde}{\widetilde}

\newcommand{\bref}[1]{\textbf{\ref{#1}}}

\renewcommand{\geq}{\,{\geqslant}\,}

\newcommand{\binner}[2]{%
  {\langle}\kern-4.15pt{\langle}#1{,}\,#2{\rangle}\kern-4.15pt{\rangle}}

\newcommand{\half}{\mathchoice{%
    \ffrac{1}{2}}{\frac{1}{2}}{\frac{1}{2}}{\frac{1}{2}}}

\newcommand{\ffrac}[2]{\raisebox{.5pt}%
  {\footnotesize$\displaystyle\frac{#1}{#2}$}\kern1pt}

\def\cA{\mathcal{A}}

\def\cI{\mathcal{I}}
\def\cJ{\mathcal{J}}

\def\cM{\mathcal{M}}

\def\cT{\mathcal{T}}

\def\cZ{\mathcal{Z}}

\numberwithin{equation}{section} \makeatletter
\@addtoreset{equation}{section}

\def\be{\begin{equation}}
\def\ee{\end{equation}}
\def\ba{\begin{array}}
\def\ea{\end{array}}

\def\dps{\displaystyle}
\def\tr{{\rm Tr}}

\newdimen\tableauside\tableauside=1.0ex
\newdimen\tableaurule\tableaurule=0.4pt
\newdimen\tableaustep
\def\phantomhrule#1{\hbox{\vbox to0pt{\hrule height\tableaurule
width#1\vss}}}
\def\phantomvrule#1{\vbox{\hbox to0pt{\vrule width\tableaurule
height#1\hss}}}
\def\sqr{\vbox{%
  \phantomhrule\tableaustep

\hbox{\phantomvrule\tableaustep\kern\tableaustep\phantomvrule\tableaustep}%
  \hbox{\vbox{\phantomhrule\tableauside}\kern-\tableaurule}}}
\def\squares#1{\hbox{\count0=#1\noindent\loop\sqr
  \advance\count0 by-1 \ifnum\count0>0\repeat}}
\def\tableau#1{\vcenter{\offinterlineskip
  \tableaustep=\tableauside\advance\tableaustep by-\tableaurule
  \kern\normallineskip\hbox
    {\kern\normallineskip\vbox
      {\gettableau#1 0 }%
     \kern\normallineskip\kern\tableaurule}%
  \kern\normallineskip\kern\tableaurule}}
\def\gettableau#1 {\ifnum#1=0\let\next=\null\else
  \squares{#1}\let\next=\gettableau\fi\next}

\def\cA{\mathcal{A}}

\def\cI{\mathcal{I}}
\def\cJ{\mathcal{J}}

\def\cM{\mathcal{M}}

\def\cT{\mathcal{T}}

\def\cZ{\mathcal{Z}}

\numberwithin{equation}{section} \makeatletter
\@addtoreset{equation}{section}

\def\ads2{\text{AdS}_{2}}
\def\cft1{\text{CFT}_{1}}

\def\be{\begin{equation}}
\def\ee{\end{equation}}
\def\ba{\begin{array}}
\def\ea{\end{array}}

\def\dps{\displaystyle}

\def\hs{\mathfrak{hs}[\lambda]}
\def\hsgl{\mathfrak{gl}[\lambda]}
\def\hsN{\mathfrak{hs}[N]}

\def\smashalg{\hsgl\rtimes{\mathbb Z}_2}
\def\trE{\mathbb{Tr}}
\def\Mat{Mat[\lambda]}
\def\MatZ{Mat[\lambda]\rtimes \mathbb{Z}_2}
\def\ehs{\mathfrak{ehs}[\lambda]}

\usepackage{jheppub}

\makeatletter
\def\@fpheader{\vspace{-.1cm}}
\makeatother

\title{On BF-type higher-spin actions in two dimensions}

\author[a,b,c]{Konstantin\ Alkalaev}

\author[d]{Xavier\ Bekaert}

\affiliation[a]{I.E. Tamm Department of Theoretical Physics, \\P.N. Lebedev Physical
Institute,\\ Leninsky ave. 53, 119991 Moscow, Russia}

\affiliation[b]{Institute for Theoretical and Mathematical Physics,\\
Lomonosov Moscow State University, \\
Leninskie Gory, GSP-1, 119991 Moscow, Russia}

\affiliation[c]{Department of General and Applied Physics, \\
Moscow Institute of Physics and Technology, \\
Institutskiy per. 7, Dolgoprudnyi, \\141700 Moscow region, Russia}

\affiliation[d]{Institut Denis Poisson, Unit\'e Mixte de Recherche 7013,\\
Universit\'e de Tours, Universit\'e d'Orl\'eans, CNRS,\\
  Parc de Grandmont, 37200 Tours, France}

\emailAdd{alkalaev@lpi.ru}
\emailAdd{xavier.bekaert@lmpt.univ-tours.fr}

\abstract{We propose a non-abelian higher-spin theory in two dimensions for an infinite multiplet of massive scalar fields and infinitely many topological higher-spin gauge fields together with their dilaton-like partners. The spectrum  includes local degrees of freedom although the field equations take the form of flatness and covariant constancy conditions because fields take values in a suitable extension of the infinite-dimensional higher-spin algebra $\hs$. The corresponding action functional is of BF-type and generalizes the known topological higher-spin Jackiw-Teitelboim gravity.
}

\vspace{7mm}

\begin{document}

\maketitle
\flushbottom

\section{Introduction}
\label{sec:candidateints}

The problem of constructing interacting theories of higher-spin (HS) gauge fields is notoriously difficult, especially at the level of the action (see e.g. \cite{Sorokin:2004ie,Bekaert:2010hw} for introductory reviews). In fact, in dimensions four and higher the examples of fully nonlinear actions compatible with the minimal coupling to the spin-two subsector are pretty scarce although such cubic interaction vertices are known since a long time \cite{Fradkin:1987ks,Fradkin:1986qy}.
On the one hand, for conformal HS gravity there exists a perturbatively local action \cite{Tseytlin:2002gz,Segal:2002gd} (see also \cite{Bekaert:2010ky}) in any even dimension, 
whose low-spin truncation gives Maxwell and Weyl actions. Unfortunately, this action expanded around conformally flat background is higher-derivative and thereby clashes with pertubative unitary.
On the other hand, nonlinear equations \cite{Vasiliev:1992av} of four-dimensional HS (super)gravity are known since several decades (and their higher-dimensional bosonic analogue \cite{Vasiliev:2003ev} since more than a decade) but it was only recently that 
action functionals were proposed \cite{Boulanger:2011dd,Boulanger:2015kfa} (see also the review \cite{Arias:2016ajh}) as an off-shell formulation of minimal bosonic four-dimensional HS gravity.
The action principles from \cite{Boulanger:2011dd,Boulanger:2015kfa} share the unusual property of being formulated in terms of differential forms on a base space of higher dimension than the spacetime manifold itself. Another example of complete HS action is  given by four-dimensional 
chiral HS gravity (i.e. the HS extension of self-dual Yang-Mills and self-dual gravity) in the light-cone formulation, both in flat \cite{Ponomarev:2016lrm,Skvortsov:2018jea} and anti de Sitter \cite{Skvortsov:2018uru} spacetimes. However, note that this action is real only in Euclidean signature.

In dimensions three and two, the situation simplifies drastically because HS gauge fields become topological. In the frame-like formulation, HS gravity theories without matter are the smooth generalizations of their spin-two counterparts. In the absence of matter, HS gauge field are described, on-shell, by a flat connection taking values in the HS algebra and, off-shell, by either a Chern-Simons (CS) action in three dimensions or by a BF action in two dimensions.
More precisely, the HS extension of CS gravity with a negative cosmological constant \cite{Achucarro:1987vz,Witten:1988hc} 
(respectively, of CS conformal gravity \cite{Horne:1988jf}) was provided in \cite{Blencowe:1988gj,Campoleoni:2010zq} (respectively, in \cite{Fradkin:1989xt,Grigoriev:2019xmp} for the conformal case) while the HS extension of Jackiw-Teitelboim gravity \cite{Teitelboim:1983ux,Jackiw:1984je,Fukuyama:1985gg}
was proposed in \cite{Alkalaev:2013fsa,Grumiller:2013swa,Alkalaev:2014qpa} (see \cite{Fradkin:1989uh,Fradkin:1989kx} for the (super)conformal case). 

Let us stress that, in three dimensions, the inclusion of matter (in the fundamental representation of the HS algebra) is known to reinstate the same level of intricacy as in four dimensions. However, in two dimensions the same degree of simplicity happens to be preserved if one includes a specific type of matter (motivated by holography):  infinitely many scalar fields with fine-tuned masses  spanning the ``twisted-adjoint'' representation of the HS algebra. 

The HS symmetry in two (or three) dimensions is identified with the Lie algebra $\hs$, originally introduced in \cite{Feigin,Vasiliev:1989re} (or, respectively, two copies thereof), where $\lambda$ is a non-negative real parameter. For integer values $\lambda=N\in\mathbb N$, this HS algebra can be truncated to the finite-dimensional algebra $\mathfrak{sl}(N,\mathbb{R})$ and the corresponding spectrum is spanned by gauge fields of spin $2,3,...,N$. However, the holographic duals of such type-N higher-spin gravity theories appear to be non-unitary conformal field theories (CFT).

In any dimension, the HS algebra completely defines the kinematics of HS gravity through several of its representations: singleton, adjoint, and twisted-adjoint. In the two-dimensional case at hand, the singleton module encodes the boundary CFT$_1$ fundamental degrees of freedom while the adjoint and twisted-adjoint modules lead, in the two-dimensional bulk, to topological HS fields and massive scalar fields respectively \cite{Alkalaev:2019xuv}. Extending the original HS algebra $\hs$, via a product with the group $\mathbb{Z}_2$ generated by an involutive automorphism (called ``twist''), allows one to unify the gravity and matter sector into a single framework. Moreover, the BF-type action associated to this extended HS algebra provides a natural extension of the HS Jackiw-Teitelboim gravity, the linearization of which reproduces the correct equations of motion for topological and local modes dictated by symmetries. This is our main result. Note that, contrary to standard BF theories, this BF-type higher-spin theory is not purely topological but includes propagating matter fields. This is possible because the gauge algebra is a subtle \textit{smash} product of an \textit{infinite-dimensional} Lie algebra with a finite group, thereby producing also fields in the \textit{twisted-adjoint} representation of $\hs$. The latter representation decomposes into infinite-dimensional irreducible representations of $\ads2$ isometry group corresponding to matter fields with local degrees of freedom. 

Prior to formulating the BF-type approach in the body of the text, it may be instructive to take a look at the problem of constructing higher-spin interactions in the metric-like formulation in order to compare the two approaches. In any dimension, there is a wide class of free gauge theories, including massless (or partially-massless) totally-symmetric fields of arbitrary spins (and depths). For each corresponding gauge theory, formally valid in any dimension $d$, one can try to set $d=2$ everywhere in the various formulae (for the action, for the gauge transformations, etc) and define ``massless'' (or ``partially-massless'') fields of given ``spin'' (and ``depth'') in two dimensions as the resulting theories. In general, one may argue that the corresponding free ``higher-spin'' gauge theories have no local degrees of freedom, but at the same time, are non-trivial off-shell, i.e. their quadratic Lagrangians are \textit{not} total derivatives (see e.g. \cite{Kessel:2018ugi}, Appendix B).\footnote{However, note the exceptional case of massless spin-$2$ fields for which the Fronsdal action is a total derivative in two dimensions, in agreement with the fact that Einstein-Hilbert action is a total derivative.}
In fact, in two dimensions the only propagating degrees of freedom are given by matter fields (either bosonic scalars or fermionic spinors) of any mass. In this sense, the issue of building consistent interactions between higher-spin gauge fields is more natural if one tries to couple them with matter fields.

The first step in this direction would be to consider Noether cubic interactions between one spin-$s$ and two spin-$0$ fields. They are of the form $\varphi_{\mu_1 ... \mu_s} J^{\mu_1 ... \mu_s}$, where $\varphi_{\mu_1 ... \mu_s}$ is the rank-$s$ metric-like gauge field and $J^{\mu_1 ... \mu_s}$ is the conserved current built out of two scalars in the standard fashion. It is well known that one can take any combination of such $(s,0,0)$-type vertices for any $s = 0,1,2,...$ without spoiling the consistency of interactions at cubic order. Note that if the Fronsdal-type theories are concerned then there are possibly also vertices of type $(s,s,0)$ and $(s,s,1)$ at $s\geq 2$ which, along with some standard lower-spin vertices, exhaust all possible types of cubic interactions in this case  \cite{Kessel:2018ugi}.\footnote{Strictly speaking, the results of \cite{Kessel:2018ugi} apply only to the flat spacetime case, but adding a cosmological constant should extend possible interaction vertices only by terms proportional to the former.} As usual, in order to go beyond the cubic order  one needs to specify the HS multiplets of both gauge and matter fields. This is where the HS algebra and its representations (see above) become crucial in the analysis of consistent interactions.

On the one hand, the scalar matter sector is not ambiguous because its spectrum is completely fixed by HS symmetries \cite{Alkalaev:2019xuv}. On the other hand, the spectrum of gauge fields to which one could try to couple this tower of scalar fields does not appear to be fixed in the metric-like formulation. For instance, one could take a sum of Fronsdal actions and start adding the above Noether cubic interactions.
This is an interesting direction that might be worth exploring (since it ensures the presence of the minimal coupling and the backreaction of matter on gravity) but we prefer to stick here to what the frame-like formulation appears to dictate. In fact, in the frame-like formulation the gauge sector is fixed, in any dimension, by the HS algebra. One should note that, in two dimensions, the corresponding metric-like formulation is slightly different from what might be expected (i.e. a sum of Fronsdal actions) from the standard scenario in higher dimensions and thus calls for some comments.
Decomposing the adjoint representation of $\mathfrak{hs}[\lambda]$ into irreducible  representations of $\ads2$ isometry algebra $\mathfrak{so}(2,1)$, one is led to define spin-$s$ gauge fields as connections which are differential 1-forms taking values in the totally-symmetric rank-$(s-1)$ representation of $\mathfrak{so}(2,1)$. In higher dimensions, such gauging of HS symmetries would generally lead us to metric-like fields of the Fronsdal type (for review, see e.g. \cite{Bekaert:2005vh}). However, in $d=2$ case the associated metric-like system can be shown to be different from the Fronsdal-type theories mentioned above \cite{Alkalaev:2013fsa,Alkalaev:2014qpa}. This is already seen in the spin-2 case without cosmological constant, where the identically-vanishing Einstein equation $G_{\mu\nu}\equiv 0$ is replaced by the Jackiw-Teitelboim equation $R=0$. Using the standard cohomological analysis of the unfolded formulation, one can show that an analogous picture is valid in the higher-spin case: the Einstein-like equation is replaced with a flatness equation. Moreover, the resulting metric-like gauge fields can be interpreted as partially-massless fields of maximal depth.\footnote{In any dimension, in the frame-like formulation, 1-form gauge fields taking values in representations of the isometry algebra described by one-row Young diagrams correspond, in the metric-like formulation, to partially-massless fields of maximal-depth \cite{Skvortsov:2006at}. The one-row Young diagrams exhaust all (independent) possibilities for $\mathfrak{so}(2,1)$, thereby suggesting the absence of a conventional frame-like formulation in two dimensions for higher-spin gauge fields which are massless (or partially-massless ones of non-maximal depth).} Note that, in any dimension, the action for the frame-like maximal-depth partially-massless fields cannot be built in terms of exterior products of the field strength 2-forms as in the Lopatin-Vasiliev type action \cite{Lopatin:1988hz}. Instead, there exists a Maxwell-like action given in terms of particular Lorentz components of the field strengths built from the gauge field \cite{Skvortsov:2006at}. In this sense, the BF action in $d=2$ is the analogue of the Lopatin-Vasiliev type action (like CS action is in $d=3$). However, the use of the BF action yields kinetic operators which are different from those obtained by taking $d=2$ in the standard metric-like theories \cite{Deser:2001us,Zinoviev:2001dt} of maximal-depth partially-massless fields. The price to pay for the BF action is that each partially-massless field comes together with a dilaton-like partner. For an infinite tower of higher-spin gauge fields, this leads to an infinite collection of dilaton-like fields. The latter collection forms a single multiplet of the higher-spin algebra (in the adjoint representation). This tower of extra fields may look unnatural from the point of view of higher-dimensional expectations, but if one looks for a higher-spin extension of Jackiw-Teitelboim gravity then this multiplet is automatically generated (by HS symmetries) from the dilaton. More generally, one may argue that the panorama of gravitational theories in two dimensions is of dilaton gravity type (see e.g. \cite{Grumiller:2002nm}). Accordingly, one may expect that their higher-spin extensions (if any) must include a dilaton-like multiplet, like HS Jackiw-Teitelboim gravity does. 

Instead of pursuing the metric-like view on the Noether program of building interactions, which is interesting on its own, our goal here is twofold: to explore how to bring together gauge and matter multiplets of the HS algebra in a unified framework inside the frame-like formulation and to initiate a program of studying their interactions at the level of BF (or Poisson Sigma model) type actions. As a first step in this direction, the present paper considers minimal coupling of matter fields to HS gauge fields via an extension of the HS algebra.

The paper is organized as follows. In Section \bref{sec:sym}, we review the HS symmetry algebra in two dimensions and its representations. In Section \bref{sec:eom}, we discuss HS-invariant equations of motion for gauge and matters fields. In Section \bref{sec:BF} the HS extension of Jackiw-Teitelboim gravity is reviewed. Section \bref{sec:ext} defines the extended HS algebra and considers the corresponding  BF-type theory. Concluding remarks are given in Section \bref{sec:con}.   

\section{Higher-spin symmetries in two dimensions}
\label{sec:sym}

The kinematics of HS gravity theories in two dimensions is entirely governed by the one-parameter family of Lie algebras $\hs$ and representations thereof. The key ingredients are (see Table \bref{tab} for a summary):

\begin{table}
\begin{center}
\begin{tabular}{|c|c|c|c|c|}
 \hline
Algebra & Infinite-dim. & Finite-dim. & Centre & Trace\\ 
(parameter) & ($\lambda\in{\mathbb R}_{\small >0}\backslash{\mathbb N}$)  & ($N\in{\mathbb N}$) &   & ($\lambda\in{\mathbb R}_{\small >0}$)\\
\hline\hline
associative & $Mat[\lambda]$ & $Mat(N,{\mathbb R})$ & $Z\cong{\mathbb R}$ & projection on $Z$ \\\hline
reductive Lie & $\mathfrak{gl}[\lambda]$ & $\mathfrak{gl}(N,{\mathbb R})$ & $\mathfrak{u}(1)$& projection on $\mathfrak{u}(1)$ \\\hline
simple Lie & $\mathfrak{hs}[\lambda]$ & $\mathfrak{sl}(N,{\mathbb R})$ & $0$ & traceless\\\hline\hline
 associative & $\MatZ$ & $Mat(N, \mathbb{R})\rtimes \mathbb{Z}_2$ & ${\cal Z}\cong{\mathbb R}$ & projection on ${\cal Z}$\\\hline
quadratic Lie & $\mathfrak{gl}[\lambda]\rtimes \mathbb{Z}_2$ & $\mathfrak{gl}(N, \mathbb{R})\rtimes \mathbb{Z}_2$ & $\mathfrak{u}(1)$ & projection on $\mathfrak{u}(1)$\\\hline
centerless Lie & $\ehs$ & $\mathfrak{esl}(N, \mathbb{R})$ & $0$ & traceless\\\hline
\end{tabular}
\end{center}
\caption{Relevant algebras}
\label{tab}
\end{table}

\paragraph{Associative vs Lie algebras.} Consider the universal enveloping algebra ${\cal U}\big(\mathfrak{so}(2,1)\big)$ of the isometry algebra $\mathfrak{so}(2,1)$ of two-dimensional anti-de Sitter spacetime $\ads2$, and its ideal 
\be
\label{ideal}
\cI = \Big({\cal C}_2-\frac14(\lambda^2-1)\Big)\,{\cal U}\big(\mathfrak{so}(2,1)\big)
\ee
generated by the eigenvalue $\frac14(\lambda^2-1)$ (for $\lambda\in\mathbb R$) of the quadratic Casimir element ${\cal C}_2\in{\cal U}\big(\mathfrak{so}(1,2)\big)$.
The quotient 
\be\label{Mlambda}
Mat[\lambda]={\cal U}\big(\mathfrak{so}(2,1)\big)/\,\cI
\ee
is an associative algebra which, for generic $\lambda\in\mathbb R$, is an infinite-dimensional analogue of the finite-dimensional associative algebra $Mat(N,{\mathbb R})$ of $N\times N$ matrices, as emphasized by our choice of notation. Moreover, for integer $\lambda=N\in \mathbb{N}$, the algebra \eqref{Mlambda} contains an infinite-dimensional ideal ${\cal J}_N$ to be factored out, and $Mat[N]/{\cal J}_N\cong Mat(N,{\mathbb R})$. The space ${\cal U}\big(\mathfrak{so}(2,1)\big)/\cI$ endowed with the commutator as Lie bracket, is a reductive\footnote{A Lie algebra is said reductive if it is a direct sum of an abelian ideal and a semisimple subalgebra.} Lie algebra, which is often denoted $\mathfrak{gl}[\lambda]$ because, for generic $\lambda\in\mathbb R$, it is an infinite-dimensional analogue of the general linear algebra $\mathfrak{gl}(N,{\mathbb R})$ \cite{Feigin}. Note that the enveloping algebra of the so-called ``Wigner deformed oscillator algebra'' provides a useful realization of $Mat[\lambda]$ \cite{Vasiliev:1989re}.

\paragraph{Higher-spin algebra.} The centre of ${\cal U}\big(\mathfrak{so}(2,1)\big)$ is spanned by the polynomials in the quadratic Casimir element ${\cal C}_2$. Accordingly, the centre of $Mat[\lambda]$ is the one-dimensional subalgebra $Z\cong\mathbb R$, which is what remains of the centre of ${\cal U}\big(\mathfrak{so}(2,1)\big)$ after quotienting the ideal \eqref{ideal}. Its Lie algebra counterpart forms a $\mathfrak{u}(1)$ ideal of $\mathfrak{gl}[\lambda]$. The Lie algebras of HS symmetries in two dimensions are traditionally defined by subtracting the one-dimensional Abelian ideal,
\be
\label{hs}
\mathfrak{gl}[\lambda]=\mathfrak{u}(1)\oplus\mathfrak{hs}[\lambda]\;,
\ee 
so that $\mathfrak{hs}[\lambda]$ is an infinite-dimensional analogue of $\mathfrak{sl}(N,{\mathbb R})$.
The structure of $\mathfrak{hs}[\lambda]$ was described in the papers \cite{Feigin,Vasiliev:1989re}, from which one may extract the following relevant facts: The Lie algebra $\mathfrak{hs}[\lambda]$ always contains $\mathfrak{so}(2,1)$ as a subalgebra. Moreover, $\mathfrak{hs}[\lambda]$ is simple if and only if $\lambda\notin\mathbb N$. The Lie algebra $\mathfrak{hs}[N]$ contains an infinite-dimensional ideal ${\cal J}_N$ to be factored out and the corresponding quotient is finite-dimensional, $\mathfrak{hs}[N]/{\cal J}_N\cong\mathfrak{sl}(N,{\mathbb R})$ \cite{Feigin,Vasiliev:1989re}. 
Consequently, the family of Lie algebras of HS symmetries in two dimensions that are simple and that allow for unitary representations, is $\mathfrak{hs}[\lambda]$ for $\lambda\notin\mathbb N$. The other algebras in the upper half of Table \bref{tab} are useful auxiliary tools (e.g. the associative algebras) or illustrative toy models (e.g. the finite-dimensional algebras) but the kinematics of pure HS gravity theories in two dimensions is determined by the one-parameter family of Lie algebras $\hs$ and representations thereof.  

\paragraph{Twist automorphism.} Let us consider basis elements $T_A = (P_a, L)$ of the Lie algebra $\mathfrak{so}(2,1)$, where $A=0,1,2$ and $a=0,1$. They have been split into transvection generators $P_a$ and Lorentz generator $L$. One can introduce the involutive automorphism $\tau$ of $\mathfrak{so}(2,1)$ acting as $\tau (P_a)=-P_a$ and $\tau(L) = L$. This automorphism can be promoted to the whole algebra ${\cal U}\big(\mathfrak{so}(2,1)\big)$ by the associativity and by setting $\tau (\mathbb{1})=\mathbb{1}$. The Casimir element ${\cal C}_2= \half T_A T^A =
P^aP_a+L^2$ is left invariant by $\tau$, therefore the automorphism $\tau$ of ${\cal U}\big(\mathfrak{so}(2,1)\big)$ descends to an automorphism of both $Mat[\lambda]$, $\mathfrak{gl}[\lambda]$ and $\mathfrak{hs}[\lambda]$, in which cases it is called ``twist'' (see e.g. \cite{Bekaert:2005vh,Alkalaev:2019xuv} for reviews). Moreover, the ideals $\cJ_N$ mentioned above are also $\tau$-invariant so that the  twist consistently descends to the finite-dimensional algebras $Mat(N, \mathbb{R})$,  $\mathfrak{gl}(N, \mathbb{R})$ and  $\mathfrak{sl}(N, \mathbb{R})$ as well. 

\paragraph{Adjoint vs twisted-adjoint representations.} Let us review two important representations of $\mathfrak{gl}[\lambda]$ on itself. Firstly, as any Lie algebra $\mathfrak{gl}[\lambda]$ acts on itself via the adjoint action, 
\be
\label{adj}
{}^*\hspace{-0.5mm}ad_y(a)=[y,a]_*:=y*a - a*y\;, \qquad \forall\, y,a\in \mathfrak{gl}[\lambda]\,,
\ee
where $*$ stands for the associative product in $Mat[\lambda]$. The same holds for its subalgebra $\mathfrak{hs}[\lambda]$.
Secondly, the twisted-adjoint action of $\mathfrak{gl}[\lambda]$ on itself is defined as 
\be\label{tadj}
{}^{\tau}\hspace{-0.5mm}ad_y(a)=[y,a]_\tau:=y*a-a*\tau(y)\,,\qquad \forall\, y,a\in \mathfrak{gl}[\lambda]\,.
\ee
Specifying $y\in \hs$ defines the twisted-adjoint action of the higher-spin algebra $\hs$ on the linear space of $\hsgl$.
 
Restricting these two actions to elements $y \in \mathfrak{so}(2,1) \subset\mathfrak{gl}[\lambda]$, we obtain the adjoint and twisted-adjoint actions of $\mathfrak{so}(2,1)$ on $\mathfrak{gl}[\lambda]$, denoted respectively as $\cT:={}^*\hspace{-0.5mm}ad_T$ and $\mathbb{T}:={}^\tau\hspace{-0.5mm}ad_T
$. The two corresponding $\mathfrak{so}(2,1)$-modules are infinite-dimensional and reducible. They can be decomposed into irreducible submodules of $\mathfrak{so}(2,1)$ which are finite-dimensional (``Killing'') modules for the adjoint action and infinite-dimensional (``Weyl'') modules for the twisted-adjoint action. The latter modules are in fact Verma modules of $\mathfrak{so}(2,1)$ with running weights expressed in terms of $\lambda$ (see \cite{Alkalaev:2019xuv} for details).

\section{Linearized higher-spin equations in two dimensions}
\label{sec:eom}

Let ${\cal M}_2$ be a two-dimensional spacetime manifold with local coordinates $x^\mu$ ($\mu=0,1$). The fields are differential $p$-forms ($p=0,1,2$) taking values in the vector space $\hsgl$. The latter will be seen, with a slight abuse of notation, as an associative algebra or as a Lie algebra depending on the context (see Table \ref{tab}). These differential $p$-forms will be denoted accordingly as $X_{[p]}\in\Omega^p({\cal M}_2)\otimes\hsgl$. 
A differential 1-form $X_{[1]}\in\Omega^1({\cal M}_2)\otimes\hsgl$ will be called a (Cartan) connection 1-form if its $\mathfrak{so}(2,1)$ piece $X^A_{[1]}T_A=e^a_{[1]}P_a+\omega_{[1]}L$ is such that the components $e^a_{[1]}$ along the transvection generators define a non-degenerate zweibein, i.e. $e_\mu^a$ is a non-degenerate $2\times2$ matrix. We will refer to this condition as the non-degeneracy condition. 

In particular, let $W_{[1]}\in \Omega^1({\cal M}_2)\otimes\mathfrak{so}(2,1)$ be an $\mathfrak{so}(2,1)$-valued connection 1-form. The respective (twisted-)adjoint covariant derivatives read
\be
\label{nabla}
\nabla  = d + W_{[1]}^A \cT_A\;, 
\qquad 
\tilde \nabla
  = d + W_{[1]}^A \mathbb{T}
	_A\;, 
\ee
where $d = dx^\mu \partial_\mu$ is the de Rham differential on $\cM_2$, while $\cT_A$ and $\mathbb{T}_A
$ are basis elements of $\mathfrak{so}(2,1)$ in the (twisted-)adjoint representations \eqref{adj} and \eqref{tadj}. 

Both squared covariant derivatives yield the curvature 2-form $R_{[2]}\in\Omega^2({\cal M}_2)\otimes\mathfrak{so}(2,1)$ as follows
\be
\label{nabla2}
\nabla^2 = R_{[2]}^A \cT_A\;,
\qquad\;\;
\tilde \nabla^2  = R_{[2]}^A\mathbb{T}_A
\;,
\ee
where 
\be
\label{zero}
R_{[2]}^A = d W_{[1]}^A + \half \epsilon^A{}_{BC} W_{[1]}^B \wedge W_{[1]}^C\;,
\ee
and  $\epsilon_{ABC}$ stands for the $\mathfrak{so}(2,1)$ Levi-Civita tensor. 

From now on, we will assume that the connection 1-form $W^A_{[1]}$ solves the zero-curvature condition $R^A_{[2]}=0$, thereby defining $\ads2$ spacetime (locally). Then, we can introduce the following covariant constancy equations
\be
\label{adj_eq}
\nabla \Omega_{[1]} \equiv (d + W_{[1]}^A \cT_A)\wedge \Omega_{[1]}  = 0 \;,
\ee 
for the adjoint-valued 1-form field $\Omega_{[1]} \in \Omega^1({\cal M}_2)\otimes\hsgl$, and 
\be
\label{twist_eq}
\tilde \nabla
 C_{[0]} \equiv (d + W_{[1]}^A \mathbb{T}_A)\, C_{[0]}  = 0\;, 
\ee 
for the twisted-adjoint-valued 0-form field $C_{[0]}\in \Omega^0({\cal M}_2)\otimes\hsgl$. 

The first equation, \eqref{adj_eq}, describes free topological HS fields that are pure gauge and thus do not carry local degrees of freedom. The second equation, \eqref{twist_eq}, describes an infinite tower of free massive scalar fields with ascending  masses \cite{Alkalaev:2019xuv}  
\be\label{Reggelike}
\left(\Box_{\ads2}+m_n^2\right)\varphi_n = 0\;, 
\qquad
m^2_n  =  \frac{(n-\lambda)(n-\lambda+1)}{R^{2}_{_{\hspace{-0.5mm}\rm AdS}}}\;,
\qquad
n=0,1,2,... \;,
\ee
where $\Box_{\ads2}$ is the wave operator on the $\ads2$ spacetime of curvature radius $R_{_{\hspace{-0.5mm}\rm AdS}}$.\footnote{A single massive scalar on constant curvature spaces is known to be described by such twisted-adjoint equations (also known as unfolded equations), see e.g. \cite{Vasiliev:1995sv}. } The space of states of each massive scalar field spans a Verma module of $\mathfrak{so}(2,1)$ with lowest energy $\Delta_n$ such that $m^2_n = \Delta_n(\Delta_n-1)$, or, equivalently, spans the particular irreducible module under the twisted-adjoint action of $\mathfrak{so}(2,1)$ discussed above.    

\section{Higher-spin Jackiw-Teitelboim gravity} 
\label{sec:BF}

The definition of a BF action requires an invariant symmetric bilinear form on the Lie algebra of symmetries. In other words, the latter algebra must be quadratic.\footnote{A symmetric bilinear form $\langle \,\,,\,\,\rangle:\mathfrak{g}\otimes\mathfrak{g}\to\mathbb{R}$ over a Lie algebra $\mathfrak{g}$ is (adjoint-)invariant if 
\be\label{quadraticLiealg}
\langle \,[a_1,a_2]\,,\,a_3\rangle=\langle a_1\,,\,[a_2,a_3]\,\rangle\,,\qquad \forall a_1,a_2,a_3\in\mathfrak{g}\,.
\ee
Lie algebras endowed with a non-degenerate invariant symmetric bilinear form are called (regular) quadratic algebras. All finite-dimensional reductive Lie algebras are quadratic. The number of independent such bilinear forms is equal to the number of abelian and simple algebras summands, e.g. two for $\mathfrak{gl}(N,{\mathbb R})=\mathfrak{u}(1)\oplus\mathfrak{sl}(N,{\mathbb R})$. It is remarkable that the \textit{infinite-dimensional} reductive Lie algebra $\mathfrak{gl}[\lambda]$ is quadratic (for generic $\lambda$).}
Fortunately, there exists a trace over $Mat[\lambda]$ realized either via deformed oscillators \cite{Vasiliev:1989re} or via the quotient algebra construction \cite{Alkalaev:2014qpa}.
By definition, it is a linear form $\tr: Mat[\lambda]\to\mathbb R$ obeying the cyclicity property 
\be\label{cyclic}
\tr\big[\,[a_1,a_2]_*\,\big]=0\,,\qquad \forall a_1,a_2\in Mat[\lambda]\,.
\ee
Equivalently, the corresponding linear form  $\tr:\mathfrak{gl}[\lambda]\to\mathbb R$ on the associated Lie algebra must be degenerate on the derived Lie algebra $\mathfrak{gl}[\lambda]^\prime$ spanned by Lie brackets.\footnote{A derived Lie algebra of a Lie algebra  $\mathfrak{g}$ is an ideal  denoted $\mathfrak{g}^\prime\subseteq \mathfrak{g}$ and defined as $\mathfrak{g}^\prime = [\mathfrak{g}, \mathfrak{g}]$. For instance, $\mathfrak{gl}(N, \mathbb{R})^\prime = [\mathfrak{gl}(N, \mathbb{R}),\mathfrak{gl}(N, \mathbb{R})]=\mathfrak{sl}(N, \mathbb{R})$ and $\mathfrak{sl}(N, \mathbb{R})^\prime = [\mathfrak{sl}(N, \mathbb{R}),\mathfrak{sl}(N, \mathbb{R})] = \mathfrak{sl}(N, \mathbb{R})$.} Note that $\mathfrak{u}(1)^\prime=0$ (since it is Abelian) and $\mathfrak{hs}[\lambda]^\prime=\mathfrak{hs}[\lambda]$ for generic $\lambda$ (since it is simple), hence $\mathfrak{gl}[\lambda]^\prime=\mathfrak{hs}[\lambda]$. Therefore, the only possibility (up to a multiplicative constant) is that the trace $\tr:\mathfrak{gl}[\lambda]\to\mathbb R$ identifies with the projector on the $\mathfrak{u}(1)$ ideal, i.e. $\tr[a]=0$ if $a\in\hs$ and $\tr[a]=a$ if $a\in\mathfrak{u}(1)$ where the centre $\mathfrak{u}(1)$ is identified with $\mathbb R$ (cf. Table \bref{tab}). Accordingly, the linear form $\tr: Mat[\lambda]\to\mathbb R$ identifies with the projector on the center $Z\cong\mathbb R$. 
An important property follows: the trace is twist-invariant, 
\be\label{twistinv}
\tr\big[\tau(a)\big]=\tr\big[a\big]\,,\qquad \forall a\in Mat[\lambda]\,,
\ee
since $\tau(a)=a$ for any $a\in\mathbb R$.

Note that the trace on $Mat[\lambda]$ automatically defines an invariant symmetric bilinear form on $\mathfrak{gl}[\lambda]$ (and on its subalgebra $\hs$ as well)
\be
\label{form}
\langle a_1\,,\,a_2\rangle_{\hsgl}\coloneqq\tr\big[a_1*a_2\big],
\ee
which is non-degenerate for non-integer $\lambda$ (see below for the case of integer $\lambda$). This parallels the relation between the Killing form on $\mathfrak{sl}(N,{\mathbb R})$ and the trace on $Mat(N,{\mathbb R})$. Note that $\mathfrak{u}(1)$ and $\mathfrak{hs}[\lambda]$ are orthogonal to each other with respect to \eqref{form} since $\tr[a_1*a_2]=a_1\tr[a_2]=0$ for $a_1\in\mathbb R$ and $a_2\in\hs$.\footnote{\label{foot}More generally, in any quadratic Lie algebra $\mathfrak{g}$ the centre $\mathfrak{z}\subset\mathfrak{g}$ and the derived algebra $\mathfrak{g}^\prime\subset\mathfrak{g}$ are the orthogonal complements of each other, as follows from the invariance condition \eqref{quadraticLiealg}. Therefore, any quadratic Lie algebra decomposes into the direct sum $\mathfrak{g}=\mathfrak{z}\oplus\mathfrak{g}^\prime$ of its center and its derived algebra (see e.g. \eqref{hs}\,).}

The HS extension \cite{Alkalaev:2013fsa,Alkalaev:2014qpa} of Jackiw-Teitelboim gravity is described in the frame-like formulation by the BF action 
\be
\label{HSJT}
S_{_{\rm{HS\,JT}}}[A,B]=\int_{{\cal M}_2} \tr[\,B_{[0]}*F_{[2]}\,]\;,
\ee
where:
\begin{itemize}
\item $B_{[0]}$ is an adjoint-valued 0-form, i.e. $B_{[0]}\in\Omega^0({\cal M}_2)\otimes\hsgl$ and $\delta_\epsilon B_{[0]}=[\epsilon_{[0]},B_{[0]}]_*$ is a gauge transformation with $0$-form gauge parameter $\epsilon_{[0]}\in\Omega^0({\cal M}_2)\otimes\hsgl$. 

\item $D = d + A_{[1]}$ is the covariant derivative of the adjoint-valued connection 1-form $A_{[1]}\in\Omega^1({\cal M}_2)\otimes\hsgl$, such that $\delta_\epsilon A_{[1]}=D\epsilon_{[0]}$ under a gauge transformation.

\item $F_{[2]} = dA_{[1]} + A_{[1]}\wedge \hspace{-0.5mm}* A_{[1]}$ is the curvature 2-form of the connection 1-form $A_{[1]}$, thus $F_{[2]}\in\Omega^2({\cal M}_2)\otimes\mathfrak{gl}[\lambda]$. Under gauge transformations, it transforms homogeneously: $\delta_\epsilon F_{[2]}=[\epsilon_{[0]},F_{[2]}]_*$.
\end{itemize}

\noindent The equations of motion following from the action \eqref{HSJT},
\be\label{BFequ}
F_{[2]}=0\;, 
\qquad
DB_{[0]}=0\;,
\ee
impose flatness and covariant constancy conditions, respectively. It is clear that $\ads2$ spacetime, i.e. \eqref{zero} with $R^A_{[2]}=0$, provides a solution of the equations \eqref{BFequ} by setting $A_{[1]}=W_{[1]}\in \Omega^1({\cal M}_2)\otimes\mathfrak{so}(2,1)$ and $B_{[0]}=0$. Let us linearize the above equations around $\ads2$ solution by setting
\be\label{linearization}
A_{[1]}=W_{[1]}+\Omega_{[1]}\,,
\ee 
i.e. $\Omega_{[1]}$ is seen as the fluctuation of the connection 1-form $A_{[1]}$ over the $\ads2$ background $W_{[1]}$.
The linearization of \eqref{BFequ} yields, respectively,
\be\label{lineareqs}
\nabla\Omega_{[1]}=0\;,
\qquad
\nabla B_{[0]}=0\;,
\ee 
where the covariant derivative $\nabla$ is defined in \eqref{nabla} (for more details see \cite{Alkalaev:2014qpa}). The first equation in \eqref{lineareqs} reproduces \eqref{adj_eq} while the second one determines the $\hsgl$-valued scalar $B_{[0]}(x)$ everywhere on the base manifold $\cM_2$ in terms of its value $B_{[0]}(x_0)$ at any given point $x_0$ (through parallel transport by the flat $\mathfrak{so}(2,1)$-connection 1-form $W_{[1]}$). These global degrees of freedom are the HS generalization of the dilaton solutions in Jackiw-Teitelboim gravity and they are in one-to-one correspondence with the (maximal-depth) Killing tensor fields of the $\ads2$ background (see \cite{Alkalaev:2014qpa} for more details).
They span an irreducible $\mathfrak{so}(2,1)$-module of dimension $2s-1$, where $s$ is the spin of the corresponding gauge field.

Two comments are in order. Firstly, the above construction of the BF action applies exactly in the same way if one replaces $\hsgl$ by $\hs$ according to \eqref{hs} everywhere in this section. Since $\mathfrak{u}(1)$ and $\mathfrak{hs}[\lambda]$ are orthogonal, the only effect is in subtracting the BF action of the $\mathfrak{u}(1)$ subsector: this  $\mathfrak{u}(1)$ term describes on-shell a constant scalar field together with a topological spin-one gauge field. 
Secondly, for $\lambda \in \mathbb{N}$ the algebra $\hs$ is not simple and contains an infinite-dimensional ideal $\cJ_N$ so that    
$\mathfrak{sl}(N, \mathbb{R}) = \hsN/\cJ_N$.
The bilinear form \eqref{form} is then degenerate \cite{Vasiliev:1989re} so that the fields taking values in the ideal $\cJ_N$  do not contribute to the action. The only non-vanishing contributions are identified with $\mathfrak{sl}(N, \mathbb{R})$-valued differential forms.    
It follows that the resulting higher-spin BF action \eqref{HSJT} then reduces to $\mathfrak{sl}(N, \mathbb{R})$ BF action \cite{Alkalaev:2013fsa}.  At $N = 2$ we reproduce the original Jackiw-Teitelboim theory in the BF form \cite{Fukuyama:1985gg}.

\section{Extended higher-spin BF-type theory} 
\label{sec:ext}

There exists an extension of the previous HS Jackiw-Teitelboim gravity where the higher-spin algebra $\hs$ is replaced with an \textit{extended HS algebra}, denoted $\ehs$, based on the trick\footnote{Note that a similar trick was also used in \cite{Vasiliev:1995sv} for a distinct proposal of two-dimensional HS gravity. The matter spectrum in \cite{Vasiliev:1995sv} is made of a single massive scalar with a fixed mass, so this proposal appears very different.} of replacing $\mathfrak{gl}[\lambda]$ by two copies of itself endowed with a subtle product between them. More precisely, the corresponding extension of $\Mat$ is defined via a smash product of the original associative algebra and a finite group $\mathbb{Z}_2$ of its automorphisms.\footnote{The smash product (also sometimes called crossed product) can be defined as follows (see e.g. the section 3.9 of \cite{smash}). Let $H$ be a group. Consider an $H$-module algebra $\cal A$ and let $\pi$ denote the corresponding action of $H$ on $\cA$. The so-called skew group ring of $H$ over $\cal A$ is denoted as ${\cal A} \# H$ and consists of pairs $(a,h)$, where $a\in \cal A$ and $h\in H$, endowed with the smash product $(a_1,h_1)\# (a_2,h_2) = \big(a_1\, \pi_{h_1}(a_2), h_1h_2\big)$, where $\pi_h$ denotes the action  of an element $h\in H$ on $\cal A$. In our case, ${\cal A} = \Mat$ and $H = \mathbb{Z}_2$, and the smash product is realized on $\Mat \# \mathbb{Z}_2$ as in \eqref{smashprod}. With a slight abuse of the standard mathematical notation, we will denote the smash product algebra as $\MatZ$. In the higher-spin theory, skew group rings of various finite groups (sometimes called outer Kleinians) over associative algebras were extensively used in constructing non-linear equations of motion, see e.g. \cite{Vasiliev:1986qx,Prokushkin:1998bq} for earlier literature and \cite{Sharapov:2018kjz,Sharapov:2019vyd} for recent studies.}

Let ${\mathbb Z}_2=\{ {\bm 1},{\bm \tau}\}$ be the group of automorphisms generated by the twist $\tau$. Then, the associative algebra denoted $Mat[\lambda]\rtimes{\mathbb Z}_2$ is the vector space spanned by elements that can be written as $a\,{\bm 1}+b\,{\bm \tau}$, where $a,b\in Mat[\lambda]$, endowed with the smash product $\star$ defined as follows
\be
\label{smashprod}
\ba{c}
(a_1\,{\bf 1}+b_1\,{\bm \tau})\star(a_2\,{\bf 1}+b_2\,{\bm \tau})=\Big(a_1*a_2+ b_1*\tau(b_2)\Big)\, {\bf 1}+\Big(a_1*b_2+ b_1*\tau(a_2)\Big)\, {\bm \tau}\,,
\ea
\ee
where $*$ denotes the product in $Mat[\lambda]$. 

The associated Lie algebra will be denoted $\mathfrak{gl}[\lambda]\rtimes{\mathbb Z}_2$. It is spanned by elements $a\,{\bf 1}+b\,{\bf {\bm \tau}}$, where $a,b\in\mathfrak{gl}[\lambda]$, endowed with the $\star$-commutator
\be
\label{starcom}
\ba{l}
[\,a_1\,{\bf 1}+b_1\,{\bm \tau}\,,\,a_2\,{\bf 1}+b_2\,{\bm \tau}\,]_\star\,=
\\
\\
\dps
\hspace{20mm}=\,\Big(\,[a_1,a_2]_*+
\big(b_1*\tau(b_2)-b_2*\tau(b_1)\big)\,\Big)\, {\bf 1} +\Big(\,[a_1,b_2]_\tau- [a_2,b_1]_\tau\,\Big) \,{\bm \tau}\,,
\ea
\ee
where $[a,b]_*=a*b-b*a$ is the $*$-commutator and $[a,b]_\tau=a*b-b*\tau(a)$ is the twisted $*$-commutator.
This implies that the $\star$-adjoint action of $\mathfrak{gl}[\lambda]$ on its extension $\mathfrak{gl}[\lambda]\rtimes{\mathbb Z}_2$ unifies the adjoint and twisted-adjoint actions of $\mathfrak{gl}[\lambda]$ on itself,
\be\label{crucialproperty}
{}^\star\hspace{-0.5mm}ad_y(a\,{\bf 1}+b\,{\bm \tau}):=[y\,{\bf 1},a\,{\bf 1}+b\,{\bm \tau}]_\star={}^*\hspace{-0.5mm}ad_y(a)\,{\bf 1}+{}^\tau\hspace{-0.5mm} ad_y(b)\,{\bm \tau}\;, \qquad \forall y,a,b\in\mathfrak{gl}[\lambda]\,,
\ee
where we used \eqref{adj}-\eqref{tadj}. 
One can show that the center  $\cZ$ of the associative algebra $\MatZ$ is the one-dimensional subalgebra $\cZ =\mathbb{R}\,{\bm 1}:= \{c\, {\bm 1}\,|\, \forall c\in \mathbb{R}\}$.

The projection to the center $\cZ$ leads to a natural trace over $Mat[\lambda]\rtimes{\mathbb Z}_2$. Let us denote by $\trE$ the trace over $Mat[\lambda]\rtimes{\mathbb Z}_2$, defined as the restriction to the trace $\tr$ over $Mat[\lambda]$, that is to say 
\be\label{defextr}
\trE[a\,{\bf 1}+b\,{\bm \tau}]=\tr[a]\;, \qquad\quad  \forall a,b\in \Mat\;.
\ee
The cyclicity property,
\be
\ba{c}
\trE\Big[\,[a_1\,{\bf 1}+b_1\,{\bm \tau},a_2\,{\bf 1}+b_2\,{\bm \tau}]_\star\,\Big]=
\tr\Big[\,[a_1,a_2]_*+
b_1*\tau(b_2)-b_2*\tau(b_1)\,\Big]\,=\,0\,,
\ea
\ee
follows from the commutation relation \eqref{starcom}, the definition \eqref{defextr}, the involution and automorphism properties $\tau^2 = \mathbb{1}$ and $\tau(a*b) =\tau(a)*\tau(b)$ of the twist, together with the properties \eqref{cyclic}, \eqref{twistinv}. Note that although the extended trace $\trE$ is degenerate along the direction of the extra generator ${\bm \tau}$, the symmetric bilinear form that it defines,
\be
\label{b_form}
\ba{l}
\dps
\langle a_1\,{\bf 1}+b_1\,{\bm \tau}\,,\,a_2\,{\bf 1}+b_2\,{\bm
\tau}\rangle_{_{\smashalg}}\coloneqq
\\
\\
\dps
\hspace{20mm}\coloneqq\trE[(a_1\,{\bf 1}+b_1\,{\bm \tau})\star(a_2\,{\bf
1}+b_2\,{\bm \tau})]
=\tr\big[a_1*a_2+ b_1*\tau(b_2)\big]\,,
\\
\\
\dps
\hspace{21mm}=\langle a_1\,,\,a_2\rangle_{\hsgl}+\langle b_1\,,\,\tau(b_2)\rangle_{\hsgl}\;,
\ea
\ee
is non-degenerate (we used \eqref{smashprod} and \eqref{form} to obtain the last two lines). The cyclicity of the extended trace and the Jacobi identity of the $\star$-commutator ensures that this symmetric bilinear form is also invariant with respect to the $\star$-adjoint action of $\mathfrak{gl}[\lambda]\rtimes{\mathbb Z}_2$ on itself. In other words, the  infinite-dimensional algebra $\mathfrak{gl}[\lambda]\rtimes{\mathbb Z}_2$ is a quadratic Lie algebra.
And these two properties (non-degeneracy and adjoint-invariance) are enough to define a proper BF action. Moreover, it also implies that this Lie algebra decomposes into the direct sum (cf footnote \ref{foot}):
\be
\label{ehs}
\smashalg = \mathfrak{u}(1) \oplus  \ehs \;, 
\ee
where the centre is the Abelian ideal $\mathfrak{u}(1)=\mathbb{R}\,{\bm 1}$ and the derived algebra, denoted $\ehs=(\smashalg)^\prime$, is the extension of the $\hs$ algebra \eqref{hs} (see Table \bref{tab}) spanned by elements $a\,{\bf 1}+b\,{\bf {\bm \tau}}$ where $a\in\hs$ and $b\in\mathfrak{gl}[\lambda]$. This extended HS algebra $\ehs$ is an ideal of $\smashalg$ and, in particular, the $\star$-adjoint action \eqref{crucialproperty} can be consistently restricted to $\ehs$.
 
A smash-product extension of the higher-spin BF-type action \eqref{HSJT} is
\be
\label{HSJT'}
S_{_{\rm{EHS\,JT}}}[{\cal A},{\cal B}]=\int_{{\cal M}_2} \trE [\,{\cal B}_{[0]}\star{\cal F}_{[2]}\,]\;,
\ee
where
\begin{itemize}
\item ${\cal B}_{[0]}=B_{[0]}\,{\bf 1}+C_{[0]}\,{\bm \tau}$ is a 0-form valued in $\smashalg$\,, with $\delta_\varepsilon{\cal B}_{[0]}=[\varepsilon_{[0]},{\cal B}_{[0]}]_\star$ as gauge transformation, where $\varepsilon_{[0]}=\epsilon_{[0]}\,{\bf 1}+\kappa_{[0]}\,{\bm \tau}$ is the gauge parameter valued in $\smashalg$.

\item ${\cal D} = d + {\cal A}_{[1]}$ is the extended HS covariant derivative of the connection 1-form ${\cal A}_{[1]}=A_{[1]}\,{\bf 1}+Z_{[1]}\,{\bm \tau}$ valued in the $\star$-adjoint representation, with $\delta_\varepsilon {\cal A}_{[1]}={\cal D}\varepsilon_{[0]}$ as gauge transformation. Note that the non-degeneracy condition (cf. the  first paragraph in Section \ref{sec:eom}) on the connection 1-form ${\cal A}_{[1]}$ is equivalent to the non-degeneracy condition of the connection 1-form $A_{[1]}$ introduced in the previous section. Let us stress there is no such condition on the 1-form $Z_{[1]}$.

\item ${\cal F}_{[2]} = d{\cal A}_{[1]} + {\cal A}_{[1]}\wedge \hspace{-0.7mm}\star \, {\cal A}_{[1]}$ is the extended curvature $\smashalg$-valued  2-form with  $\delta_\varepsilon {\cal F}_{[2]}=[\varepsilon_{[0]},{\cal F}_{[2]}]_\star$ as gauge transformation.
\end{itemize}

The equations of motion following from the BF-type action \eqref{HSJT'} impose the on-shell flatness and covariant constancy conditions 
\be
\label{extHSJTeom}
{\cal F}_{[2]}=0\;, 
\qquad
{\cal D}{\cal B}_{[0]}=0\;,
\ee
which are natural extensions of \eqref{BFequ}. In components, using \eqref{smashprod} and \eqref{starcom} we obtain 
\be
\label{F1}
F_{[2]} + Z_{[1]} \wedge *\, \tau(Z_{[1]}) = 0\;,  
\ee
\be
\label{Ftau}
dZ_{[1]} + A_{[1]}\wedge *\, Z_{[1]} + Z_{[1]} \wedge *\, \tau(A_{[1]}) = 0\;,  
\ee
and 
\be
\label{DB1}
DB_{[0]}  + Z_{[1]}* \tau(C_{[0]}) - C_{[0]}* \tau(Z_{[1]}) = 0\;,  
\ee

\vspace{-2mm}

\be
\label{DBtau}
\tilde{D}C_{[0]} -  [\,B_{[0]}, Z_{[1]}\,]_\tau = 0\;,  
\ee
where we used $F_{[2]} = dA_{[1]} + A_{[1]}\wedge * A_{[1]}$ and $D = d+ [A_{[1]}, \,\cdot\;]_*$ which are respectively the $\hsgl$ curvature and the $*$-adjoint covariant derivative (as defined in Section \bref{sec:BF}) and also introduced the twisted-adjoint covariant derivative $\tilde{D} = d + [A_{[1]}, \, \cdot \;]_\tau$\,. In this way, we emphasize that $\smashalg$ BF-type theory contains $\hsgl$ BF standard theory as a subsector provided  that all fields are truncated to 
$\hsgl\subset \smashalg$, i.e. 
\be
\label{HSBF}
{\cal A}_{[1]}=A_{[1]}\,{\bf 1}\;, 
\qquad 
{\cal B}_{[0]}=B_{[0]}\,{\bf 1}\;.
\ee 
Thus, the extended HS action \eqref{HSJT'} can be reduced to the HS Jackiw-Teitelboim action \eqref{HSJT} through the truncation \eqref{HSBF}. 

We note that, upon imposing an appropriate gauge choice on-shell, the 1-form $Z_{[1]}$ can be set to zero.  Indeed, the equation \eqref{Ftau} implies that, locally, $Z_{[1]}=\tilde{D}K_{[0]}$ for some 0-form $K_{[0]}$ provided  $A_{[1]}$ satisfies the equation \eqref{F1}. Then, writing down the component form of the gauge transformations in the 1-form sector,
\be
\label{deltaA}
\delta A_{[1]} =D\epsilon_{[0]}+ Z_{[1]}* \tau(\kappa_{[0]}) - \kappa_{[0]}* \tau(Z_{[1]}) \;,  
\ee
\be
\label{deltaZ}
\delta Z_{[1]} =\tilde{D}\kappa_{[0]} -  [\,\epsilon_{[0]}, Z_{[1]}\,]_\tau \;,  
\ee
we can see that making use of the gauge parameter $\kappa_{[0]}$  the field $Z_{[1]}$ can be set to zero, on-shell. Such a gauge condition does not constrain the other  gauge parameter $\epsilon_{[0]}$\,, since $[\epsilon_{[0]}, Z_{[1]}]_\tau =0$ if $Z_{[1]}=0$. In particular, we are left with $\delta_\epsilon A_{[1]} =D\epsilon_{[0]}$ for any $\epsilon_{[0]}\in\Omega^0({\cal M}_2)\otimes\hsgl$. Note that such the gauge condition $Z_{[1]}=0$ is accessible, except if one imposes by hand a non-degeneracy condition on $Z_{[1]}$  similar to the one on $A_{[1]}$. However, we do not see presently  any clear motivation for introducing such a second zweibein-like field, so this partial gauge is indeed accessible here. When $Z_{[1]}=0$, note that the equation \eqref{F1} reduces to $F_{[2]}=0$.

To summarize, the flatness condition for the 1-form $\cA_{[1]}$ in \eqref{extHSJTeom} implies that, locally, the extended 1-form connection describes $\ads2$, i.e. ${\cal A}_{[1]}=W_{[1]}\,{\bf 1}$ in a suitable gauge.  Then, due to the property \eqref{crucialproperty}, the covariant constancy condition for the 0-form  ${\cal B}_{[0]}$ \eqref{extHSJTeom} decomposes as 
\be
{\nabla}B_{[0]}=0\;, 
\qquad 
\tilde{\nabla}
C_{[0]}=0\;.
\ee 
In this sense, the equations of motion \eqref{HSJT'} of the extended BF-type theory  can be thought of as a higher-spin covariantization of the twisted-adjoint equation \eqref{twist_eq} on the extra 0-form $C_{[0]}$, together with the adjoint equation \eqref{lineareqs} on the 0-form $B_{[0]}$ already present in the HS Jackiw-Teiteboim theory. 

Let us consider another consistent truncation of the equations of motion \eqref{extHSJTeom}, 
\be
\label{forth}
{\cal A}_{[1]}=A_{[1]}\,{\bf 1}\;,
\qquad
{\cal B}_{[0]}=C_{[0]}\,{\bm \tau}\;,
\ee 
that leads to the system  
\be
F_{[2]}=0\;,
\qquad 
\tilde{D}C_{[0]}=0\;.
\ee
The linearization of these equations around $\ads2$ solution $W_{[1]}$ \eqref{zero} via the decomposition \eqref{linearization} of the connection 0-form $A_{[1]}$ are respectively 
\be
\label{eqs_fin}
\nabla\Omega_{[1]}=0\;, 
\qquad
\tilde{\nabla}
C_{[0]}=0\;,
\ee 
which perfectly reproduces the linear equations \eqref{adj_eq} and \eqref{twist_eq}.

Let us add a few summarising remarks on the field content, both off-shell and on-shell. Off-shell we have the following fields:

\begin{enumerate}

\item  The field $A_{[1]}$ is the usual HS connection, on-shell it describes locally the $AdS_2$ solution $W_{[1]}$ (as it does for spin-two Jackiw-Teitelboim gravity) because $A_{[1]}$ is non-degenerate.

\item The field  $B_{[0]}$ is the HS extension of the dilaton of Jackiw-Teitelboim gravity. It plays the role of Lagrange multiplier for the flatness equation on $A_{[1]}$ and is thereby crucial in order to have an action principle for exactly the same reason as in the spin-two case (for which the Einstein-Hilbert action is a pure boundary term in two dimensions, hence spin-two gravity theories in two dimensions require the addition of a dilaton field). On-shell, this 0-form will describe the infinite collection of maximal-depth Killing tensor fields for all spins. For instance, the dilaton in Jackiw-Teitelboim-gravity carries the fundamental representation of $\mathfrak{so}(2,1)$. More generally, these 0-forms are topological in the sense that each of them is described on-shell by a finite-dimensional representation of $\mathfrak{so}(2,1)$.

\item The field $Z_{[1]}$ can be thought as the Lagrange multiplier for the twisted-adjoint equation but its true role is more than that.\footnote{Strictly speaking, the field $Z_{[1]}$ is \textit{not} a mere Lagrange multiplier since the action contains a term (necessary for gauge invariance) which is quadratic in $Z_{[1]}$ (and linear in $B_{[0]}$).} In fact, in any dimension $d$ the twisted-adjoint equation (alone) can be obtained from a BF-type action principle by introducing a Lagrange multipler which is a $(d-1)$-form taking values in the same representation. In general, the question is then whether this additional set of fields introduces new degrees of freedom and whether it fits in a natural way within the remaining set of fields. In the present case, the field $Z_{[1]}$ is pure gauge on-shell (distinctly from $A_{[1]}$ which is topological but not pure gauge, since $A_{[1]}$ is non-degenerate while $Z_{[1]}$ is degenerate) and it allows to make gauge symmetry manifest with respect to the extended HS algebra that unifies the gauge and matter sector. Note that, although $Z_{[1]}$ is pure gauge on-shell, it is not pure gauge off-shell and actually plays a crucial role in the action.

\item The field $C_{[0]}$ is the dynamical field, since this 0-form carries on-shell the local degrees of freedom of an infinite tower of scalar fields with fine-tuned mass.

\end{enumerate}

\noindent The on-shell physical spectrum is as follows: 

\begin{enumerate}

\item[] \textit{Topological sector}: an infinite tower of dilaton-like scalar fields $\chi^{(s)}$, describing as a whole the adjoint representation of the higher-spin algebra $\hsgl$ and decomposing under $\mathfrak{so}(2,1)$ into the infinite collection of maximal-depth Killing tensors for the gauge fields of spin $s=1,2,3,...\,$. Each dilaton-like field carries a finite-dimensional representation of $\mathfrak{so}(2,1)$.

\item[] \textit{Dynamical sector}: an infinite tower of dynamical scalar fields $\varphi_n$, describing as a whole the twisted-adjoint representation of the higher-spin algebra $\hsgl$ and decomposing under $\mathfrak{so}(2,1)$ into the infinite collection of massive scalars satisfying \eqref{Reggelike} for $n=0,1,2,...$ Each scalar field carries an infinite-dimensional representation of $\mathfrak{so}(2,1)$.

\end{enumerate}

Finally, let us make two comments similar to the ones at the end of the previous section. First, to construct the BF-type action one may consider differential forms taking values in the subalgebra $\ehs$ (instead of $\smashalg$) in order to get rid of the $\mathfrak{u}(1)$ subsector. More explicitly, this means that $A_{[1]}$ and $B_{[0]}$ are restricted to $\hs$, as in HS Jackiw-Teitelboim gravity, while $Z_{[1]}$ and $C_{[0]}$ still take values in the whole algebra $\hsgl$.
For instance, in the truncation \eqref{forth} the $\mathfrak{u}(1)$-connection decouples from the adjoint sector, while scalar components are still present in the twisted-adjoint sector that keeps the equations \eqref{eqs_fin} consistent.  
Second, we note that at integer values of the parameter (i.e. $\lambda =N\in \mathbb{N}$), the extended HS algebra $\mathfrak{gl}[N]\rtimes{\mathbb Z}_2$ contains an infinite-dimensional ideal denoted $\cJ_N\rtimes \mathbb{Z}_2$, where $\cJ_N$ are infinite-dimensional ideals in $\mathfrak{gl}[N]$ (see Section \bref{sec:BF}). The resulting quotient 
\be
\label{ext_fin}
\frac{\mathfrak{gl}[N]\rtimes{\mathbb Z}_2}{\cJ_N \rtimes \mathbb{Z}_2} = \{a \,{\bm 1} + b\,{\bm \tau}\,|\,\, \forall \,a,b \in \mathfrak{gl}(N, \mathbb{R})\}\;
\ee
is a finite-dimensional Lie algebra which will be denoted as $\mathfrak{gl}(N, \mathbb{R})\rtimes \mathbb{Z}_2$. Its centerless part is defined from the decomposition (see Table \bref{tab}):\footnote{Our notation $\mathfrak{esl}(N,\mathbb{R})$ is introduced  by analogy with the notation $\mathfrak{ehs}[\lambda]$ for the infinite-dimensional extended HS algebra. It simply means that $\mathfrak{esl}(N,\mathbb{R})$ is an extension of $\mathfrak{sl}(N,\mathbb{R})$ obtained by factoring out the $\mathfrak{u}(1)$ center from the quotient $\mathfrak{gl}(N, \mathbb{R})\rtimes \mathbb{Z}_2$. It would be interesting to study its structure to clarify whether it is (semi)simple or not. An explicit realization of the twist on $\mathfrak{gl}(N, \mathbb{R})$ and respective equations will be considered elsewhere.}
\be
\mathfrak{gl}(N, \mathbb{R})\rtimes \mathbb{Z}_2=\mathfrak{u}(1)\oplus\mathfrak{esl}(N, \mathbb{R})\,.
\ee 
One may consider a BF theory with the gauge algebra $\mathfrak{gl}(N, \mathbb{R})\rtimes \mathbb{Z}_2$ which yields a topological system of equations of motion which are (twisted-)adjoint covariant constancy conditions on finite-dimensional field spaces. In particular, for  \eqref{forth} the equations of motion are reduced to the standard $\mathfrak{gl}(N, \mathbb{R})$ BF equations along with new topological equations in the sector of $0$-forms. The latter are analogous to those discussed in three-dimensional HS theory \cite{Barabanshchikov:1996mc,Prokushkin:1998bq} as a topological subsystem decoupled from the original dynamical  twisted-adjoint equations at integer $\lambda=N$.

\section{Concluding remarks}
\label{sec:con}

To summarize, the non-Abelian BF-type action \eqref{HSJT'} provides a natural extension of the action \eqref{HSJT} of HS Jackiw-Teitelboim gravity via the addition of a matter multiplet. The corresponding equations of motion describe  an infinite tower of scalar fields with fine-tuned increasing masses \eqref{Reggelike},  coupled to the topological gauge fields of HS Jackiw-Teitelboim gravity. In particular, their linearization around $\ads2$ background reproduces the correct equations fixed by the HS symmetries in two dimensions. Let us stress that our construction of the extended HS gravity action \eqref{HSJT'} with the above properties relies only on the existence of a twist-invariant trace on the HS algebra, not on its explicit form (though the latter would become important to write down the expression of the action in components). 

The existence of the BF-type action \eqref{HSJT'} is remarkable and contrasts with the situation in three-dimensional HS gravity where the inclusion of matter in the fundamental representation of the HS algebra in three dimensions (hence in the ``bifundamental'' representation of $\hs$, since the latter algebra comes in two copies) is known to reinstate the same level of intricacy as in dimension four.\footnote{For instance, the two action principles \cite{Prokushkin:1999gc,Bonezzi:2015igv} which have been proposed for the nonlinear equations of motion in \cite{Prokushkin:1998bq} are not usual CS actions (in particular, the action in \cite{Prokushkin:1999gc} has no base space while the base space of the action in \cite{Bonezzi:2015igv} is of higher dimension than the spacetime manifold).}

The nonlinear equations \eqref{extHSJTeom} are the two-dimensional analogues of the equations considered in the approach of \cite{Sharapov:2019vyd} (and references  therein) to the Noether procedure in the unfolded formulations of higher-dimensional HS gravity theories. More precisely,  following the same logic as \cite{Sharapov:2019vyd}, applied to two dimensions, one should consider a \textit{deformation} of the extended HS algebra considered above.\footnote{We thank A. Sharapov and E. Skvortsov for discussions on this point.} Note that if this deformed extended HS algebra admits a trace, then the present BF-type construction would generalize to this deformed case as well.

BF-type HS  theories in two dimensions obviously require further study. In particular, the physical content of the untruncated spectrum in the above model should be analyzed further for several reasons. For instance, there seems to be no genuine non-linearity in the matter fields (the field equations are linear in the 0-forms) nor backreaction on the gauge fields from the presence of matter (the 0-form sector does not source the 1-form sector). This feature might be improved by making use of the interactions generated from a deformation of the extended HS algebra. Let us point out that, since any BF-type action takes the form of a topological-like\footnote{Note that here the target space is the dual of the extended HS algebra. The latter involves a smash product of an infinite-dimensional algebra. This is the reason why this BF-type action can (and does) describe local degrees of freedom.} Poisson Sigma model, a reasonable expectation is that the fully interacting action still takes the form of a Poisson Sigma model, whose Poisson bivector field is a nonlinear deformation of the undeformed linear one. Last but not least, for holography one should add some right boundary terms and check whether the corresponding total action may capture correlators of single-trace operators in some suitable CFT$_1$.

\vspace{3mm}
\noindent \textbf{Acknowledgements.} We are grateful to M. Grigoriev, K. Mkrtchyan, A. Sharapov, E. Skvortsov, M. Vasiliev for useful discussions. The work of K.A. was supported by the Russian Science Foundation grant 18-72-10123.


\providecommand{\href}[2]{#2}\begingroup\raggedright\endgroup

\end{document}